\documentclass{mnras}
\usepackage[T1]{fontenc}
\usepackage{ae,aecompl}
\usepackage{graphicx}	
\usepackage{amsmath}	
\usepackage{amssymb}	
\DeclareGraphicsExtensions{.ps}

\title{Results from the SuperModel Analysis of the X-COP Galaxy Clusters Sample}
\author[R.Fusco-Femiano]{
Roberto Fusco-Femiano$^{1}$\thanks{E-mail: roberto.fuscofemiano@iaps.inaf.it}
\\
$^{1}$IAPS-INAF, via del Fosso del Cavaliere, 00133 Roma, Italy}

\date{Accepted XXX. Received YYY; in original form ZZZ}

\pubyear{2015}

\begin{document}
\label{firstpage}
\pagerange{\pageref{firstpage}--\pageref{lastpage}}
\maketitle

\begin{abstract}
In this work, the analysis of the stacked projected temperature profile of twelve galaxy clusters of the X-COP sample is performed via the SuperModel, a tool for investigating the intracluster medium thermodynamic properties already tested on many clusters since 2009. Two separate fits have been carried out: to the joint X-ray and SZ projected temperature profiles and to the X-ray data only. The entropy, thermal pressure and hydrostatic mass profiles of the cluster sample and the level of the nonthermal pressure component in the cluster outskirts are determined combining the deprojected temperature profile with the stacked electron gas density profile corrected for clumpiness. The first analysis gives results in good agreement with those reported in Ghirardini et al. (2018): a modest presence of the nonthermal support and an entropy profile that follows at distances beyond $r_{500}$ the predicted power law increase with slope 1.1. On the contrary, the results of the second analysis are consistent with the findings obtained by the \textit{Suzaku} observations, steep temperature and flat entropy profiles, with a higher level of the nonthermal pressure component in agreement with the values derived by numerical simulations. The conclusion is that a steep temperature profile could be present in the outskirts of the X-COP cluster sample instead of the flatter temperature profile reported by SZ observations.
\end{abstract}

\section{Introduction}

Since more than one decade the study of clusters of galaxies is mainly addressed to their outskirts. In fact, substantial amounts of the baryons and of the dominant dark matter (DM) component reside in these external regions. Besides, connecting the intracluster medium (ICM) with the surrounding 
environment, the cluster outskirts are sites of physical processes and events (see Kravtsov \& Borgani 2012; Cavaliere \& Lapi 2013; Reiprich er al. 2013; Walker et al. 2018).

The formation of large scale structure generates bulk and turbulent motions that give raise to a nonthermal pressure component  that, according to hydrodynamical simulations, increases going toward the
outskirts of galaxy clusters (Vazza et al. 2009; Valdarnini 2011; Lau et al. 2013; Gaspari \& Churazov 2013; Nelson et al. 2014). Quantifying the level of this nonthermal pressure is fundamental for our understanding of the
ICM thermodynamical properties, that give remarkable information on the galaxy cluster formation. Furthermore, it also required to
derive the bias in the mass estimation when only the assumption of thermal hydrostatic equilibrium (HE) is considered. This in turn may affect the determination of cosmological parameters from
clusters of galaxies.

The capability of investigating the ICM in the cluster outskirts via bremsstrahlung emission in X rays has been severely limited, until recently, by the low density of these regions. A significant breakthrough in this investigation has been obtained by the advent of the \textit{Suzaku} X-ray observatory. With its low and stable particle background, it has made possible the spectroscopic X-ray study of these external regions obtaining somewhat unexpected findings observed in several clusters (Akamatsu et al. 2011; Reiprich et al. 2013; Walker et al. 2013). Namely, a rapid decline of the ICM temperature often accompanied by an unphysical decrease of the X-ray mass
profile. This leads to a gas mass fraction $f_{gas}$ higher than the cosmic value (Simionescu et al. 2011; Fusco-Femiano \& Lapi 2013, 2014); an entropy flattening beyond $r_{500}$ with respect to the power law increase with slope $1.1$ expected from external gas accretion under pure gravitational infall (Tozzi \& Norman 2001; Voit 2005; Lapi et al. 2005; Borgani et al. 2005; Lau 2015); azimuthal variations of the ICM thermodynamical properties with a more efficient thermalization observed in infall regions overlooking filamentary structures of the cosmic web (Kawaharada et al. 2010; Ichikawa et al. 2013; Sato et al. 2014).

Several mechanisms have been proposed to explain these findings, such as gas clumping, different time-scales for thermalization of ions and electrons, breakdown of the HE due to the presence of turbulence and bulk motions, weakening of the accretion shocks
due to a slowdown in the growth rate at late cosmic times, or from some still unknown systematics that lead to a rapid decline of the temperature in the cluster outskirts observed by \textit{Suzaku} (e.g., see Fusco-Femiano \& Lapi 2015; Walker et al. 2018).

A complementary investigation of the ICM in the cluster outskirts is given by the SZ effect due to
the inverse Compton scattering of the cosmic microwave background (CMB) photons by the hot ICM electrons that
produces a spectral shift of the CMB detection. This effect depends on the ICM thermal pressure integrated along the line of sight
(Sunyaev \& Zeldovich 1972). The temperature profile is derived from the SZ pressure profile, $P_{SZ}$, combined with the X-ray electron density profile ($T_{SZ} = P_{SZ}/n_e$).

The analysis of 12 galaxy clusters of the XMM Cluster Outskirts Project (X-COP) (Eckert et al. 2017), the same sample used by Ghirardini et al. (2018; hereafter G18) to investigate the ICM thermodynamic properties out to the virial radius, is performed here with the SuperModel (SM, Cavaliere et al. 2009).
The SM is a semi-analytic tool to investigate the ICM thermodynamic properties
based on few physical parameters of the underlying entropy state of the intracluster medium. Since 2009, several clusters have been successfully analyzed with the SuperModel (Fusco-Femiano, Cavaliere \& Lapi 2009; Fusco-Femiano et al. 2011), and more recently it has been used to highlight the role of the nonthermal pressure component in the cluster outskirts (Fusco-Femiano \& Lapi 2013, 2014, 2015, 2018).

The X-COP sample contains eight non-cool-core (NCC) clusters and
four cool-core (CC) systems. The stacked temperature and electron density
profiles of the sample (see G18) are considered in two separate SM analyses. The first considers the projected temperature profile obtained by
combining the \textit{XMM-Newton} X-ray data observed up to $r_{500}$ with the SZ data derived by the \textit{Planck} survey beyond $r_{200}$, while the second regards the X-ray data only.

The paper is organized as follows. In the next Section 2, the stacked temperature profile of the X-COP sample given by \textit{XMM-Newton} observations and by the $T_{SZ}$ profile is analyzed. Then the analysis of only the X-ray temperature profile. The results and the conclusions are drawn in Section 3. The SM equations are reported in Appendix A, yielding the temperature, pressure and total mass also when a nonthermal pressure component is included in the HE. The temperature, pressure and entropy profiles are normalized to their median values at $r_{500}$ reported in Section 2.

Throughout the paper the standard flat cosmology is adopted with parameters in
round numbers: $H_0 = 70$ km s$^{-1}$ Mpc$^{-1}$, $\Omega_{\Lambda} = 0.7$,
$\Omega_M = 0.3$ (\textit{Planck} collaboration XIII 2016).

\section{SuperModel analysis of the X-COP clusters sample}

As reported in the Introduction, two separate SM analyses are performed considering in the first the stacked temperature profile obtained by adding to the \textit{XMM-Newton} X-ray temperature data the SZ temperature profile derived by the \textit{Planck} survey, investigating in such way a cluster region that extends beyond $r_{200}$; in the second only the X-ray data measured up to $r_{500}$ are considered. The gas density profile is traced by Ghirardini et al. (2018) applying two different deprojection methods to the X-ray surface brightness finding a good agreement between them. The X-ray electron density profile $n_e$ has been corrected for the presence of clumpiness applying the azimuthal median method reported in Eckert et al. (2015). The presence of inhomogeneities in the accreted gas is expected to be relevant for the X-ray flux measured beyond $r_{500}$ (see Vazza et al 2013; Roncarelli et al 2013;
Eckert et al. 2015). This method allows to
disentangle the effects of gas clumping from the possible presence of a nonthermal pressure component in the outskirts of galaxy
clusters.

The analysis of the temperature profiles is realized with the entropy-based SM derived by the HE equation when the entropy distribution
$k = k_B T/n^{2/3}$ is specified. The assumed shape is $k(r) = k_c + (k_R - k_c)(r/R)^a$ (see Voit 2005) where $k_c$ is the central entropy level set by feedback from astrophysical sources and radiative cooling; $k_R$ is the entropy at the virial radius $R$ produced by supersonic gas inflows from the surrounding environment into the DM gravitational potential well; $a$ is the slope of the power law increase from $k_c$ (see Cavaliere et al. 2009 for more details).
To satisfy the steep temperature and flat entropy profiles observed by \textit{Suzaku} toward the virial radius in several CC clusters, and in the directions of NCC ones not disturbed by mergers (like Coma, Simionescu et al. 2013), Lapi, Fusco-Femiano \& Cavaliere (2010) have considered a modified entropy shape for the SM that starts as a power law with slope $a$, but flattens at distances greater than $r_b$. For the sake of simplicity, a linear entropy decline with gradient $a^{\prime}\equiv (a - a_R)/(R/r_b -1)$ is assumed, where $r_b$ and $a^{\prime}$ are free parameters to be determined from the fitting of the temperature profile.

The stacked profiles of the electron density, pressure, temperature and entropy of the X-COP galaxy clusters sample used in the SM analyses are reported in Table D.2 of G18. In particular, the thermodynamical profiles of pressure, temperature and entropy are scaled with the self-similar quantities at an overdensity of 500. The scale radii and the self-similar scaling quantities are estimated using the mass values reported by Ettori et al. (2018). Their values of $r_{500}$ give a median value of 1256 Kpc, and for
the self-similar quantities median values of $2.68 \times 10^{-3}$ keV cm$^{-3}$ for $P_{500}$, 6.50 keV for $T_{500}$ and
1174.1 kev cm$^{-2}$ for $K_{500}$ are derived by Eq.s 8, 10 and 12 in G18, respectively.

\subsection{SM analysis of the \textit{XMM-Newton} and \textit{Planck} temperature data}

The first analysis regards the temperature profile (see Fig. 1) obtained by adding to the \textit{XMM-Newton} data the three SZ temperature points at $r > r_{500}$ ($T_{SZ} = P_{SZ}/n_e$); the SZ point at $r = 0.53r_{500}$ appears to be inconsistent with the X-ray value at the same distance. Just by looking Fig. 1, it is evident that the temperature gradient shown by only the X-ray data is steeper than that given by the X-SZ temperature points.
For this profile the SM fit requires an entropy power law increase toward $R$ with slope $a$ ($k(r) = k_c + (k_R - k_c)(r/R)^a$). The SM temperature profile used in this analysis is given by Eq. A1  when only the thermal pressure, $p_{th}$, supports the hydrostatic equilibrium (HE).
\begin{figure}
\includegraphics[width=\columnwidth]{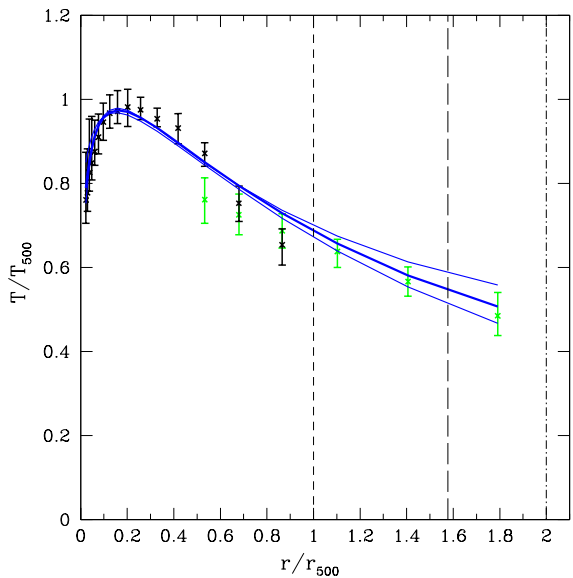}
\caption{The black points indicate the stacked \textit{XMM-Newton} projected temperature profile, while the green points are the stacked projected temperature data $T_{SZ}$. The thick blue line shows the SM joint fit to the X-ray and SZ data 
($\chi^2_{red}$ = 9.33/15); the thin lines represent the $1\sigma$ error. The vertical dashed, long-dashed and dot-dashed lines represent $r_{500}$, $r_{200}$, and the virial radius $R$, respectively.}
\label{fig:xcop_temp_XSZ_figure}
\end{figure}
This implies $\delta(r) \equiv p_{nth}/p_{th}$ = 0, where $p_{nth}$ is the nonthermal pressure component. The assumed functional shape, $\delta(r)$, in agreement with the indication of numerical simulations (e.g., Lau et al. 2009; Vazza et al. 2011), is given by:

\begin{equation}
\delta(r) = \delta_R\, e^{-(R-r)^2/l^2}
\end{equation}
which decays on the scale $l$ toward the inside from a round maximum (see Cavaliere et al. 2011). The virial radius $R$ is assumed to be $2r_{500}$.

The gas mass fraction $f_{gas} \equiv M_{gas}/M$ profile of Fig. 2 (dashed area) is obtained by combining the deprojected SM temperature profile, $T_{X-SZ}^{SM}$, with the electron gas density profile corrected for clumpiness. The gas mass fraction results at $R$ higher than the universal value highlighting the presence of a
nonthermal pressure component necessary for the hydrostatic equilibrium. To evaluate the level of this component, a new fit to the temperature profile with $\delta_R > 0$ in Eq. A1 is performed until a greater total mass $M$ is obtained that gives a value
of $f_{gas}$ in agreement with the universal value at the virial radius (black line in Fig. 2). A modest level of this component with respect to the
total pressure $p_{tot}$ is derived, namely $\alpha \equiv p_{nth}/p_{tot} \sim (10-15)\%$ at $R$ that reduces to $\sim(9-13)\%$ at $r_{200}$ and to $\sim(5-11)\%$ to $r_{500}$ according to Eq. 1, with $\delta_R$ in the range (0.11-0.18) and $l = 0.6$ that gives the best profile that satisfies the universal gas mass fraction. For $\delta_R > 0$ the best fit temperature profile is practically 
coincident with the profile of Fig. 1 for the modest level of $\alpha(R)$. 
These levels of turbulence are consistent with the median values derived by Eckert et al. (2018) for the same sample: $\alpha(r_{200}) = 10.5^{+4.3}_{-5.5}\%$ and $\alpha (r_{500}) = 5.9^{+2.9}_{-3.3}\%$.

\begin{figure*}
\begin{center}
\parbox{16cm}{
\includegraphics[width=0.4\textwidth,height=0.3\textheight,angle=0]{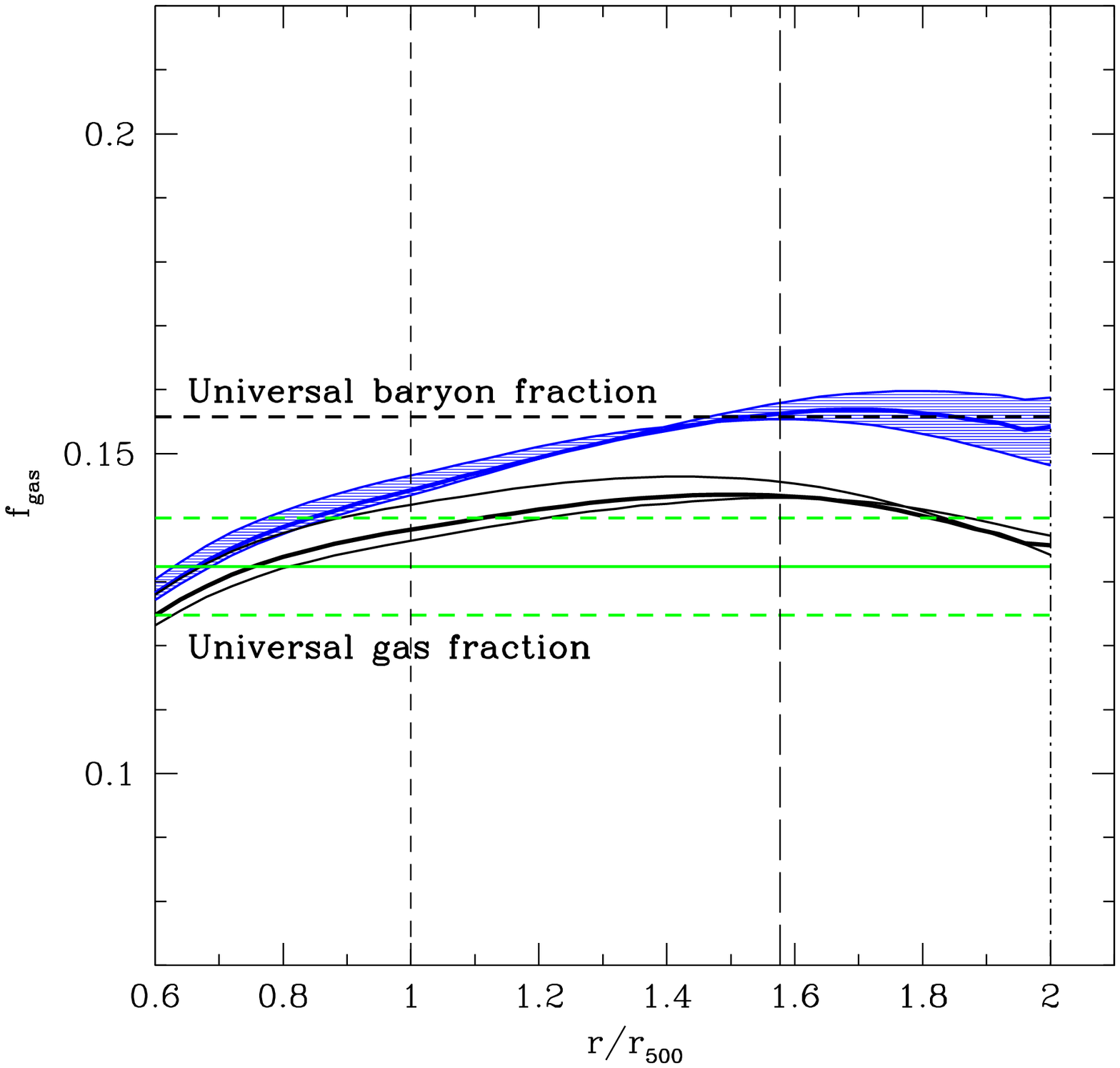}
\includegraphics[width=0.4\textwidth,height=0.3\textheight,angle=0]{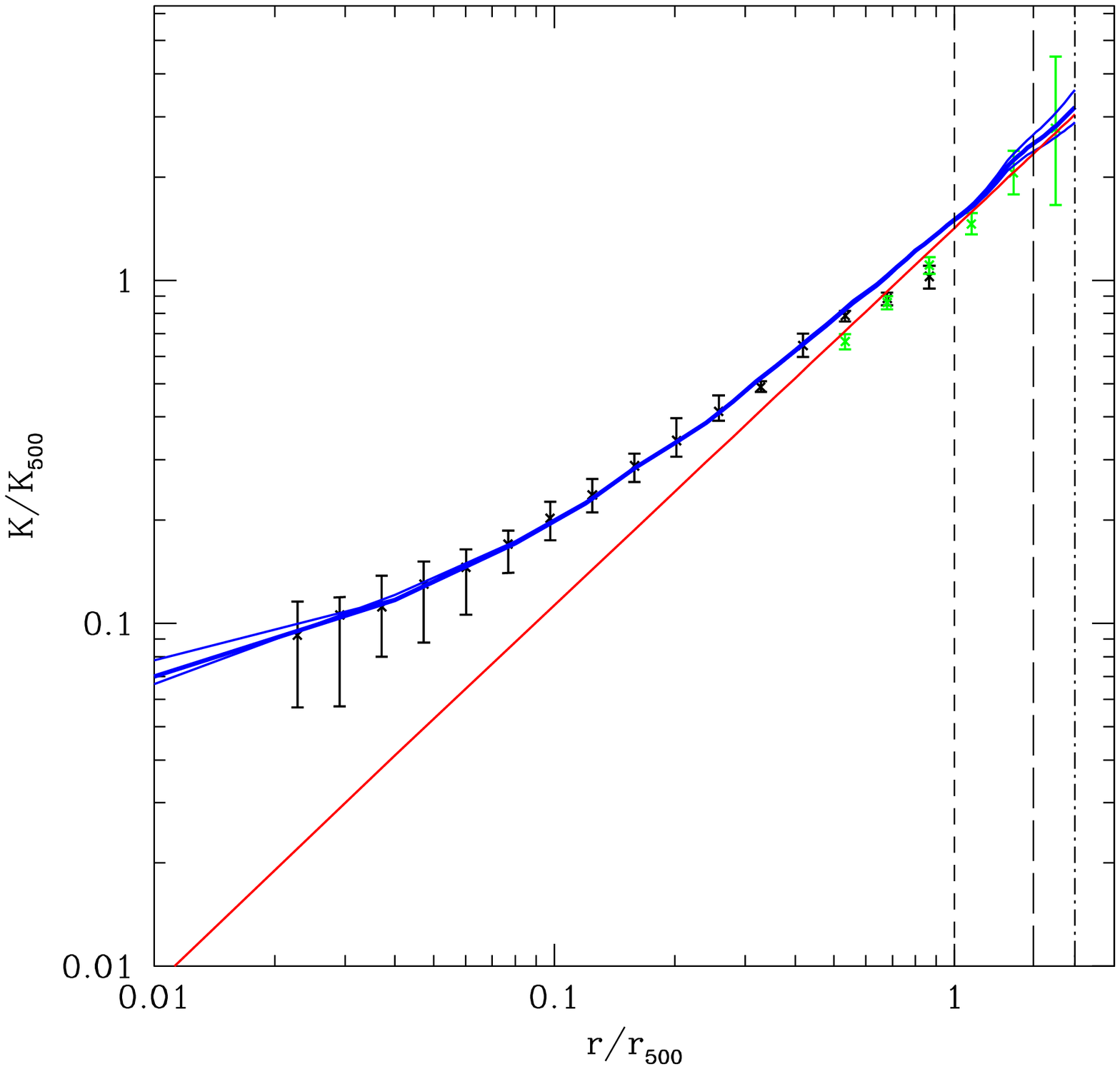}}
\parbox{16cm}{
\includegraphics[width=0.4\textwidth,height=0.3\textheight,angle=0]{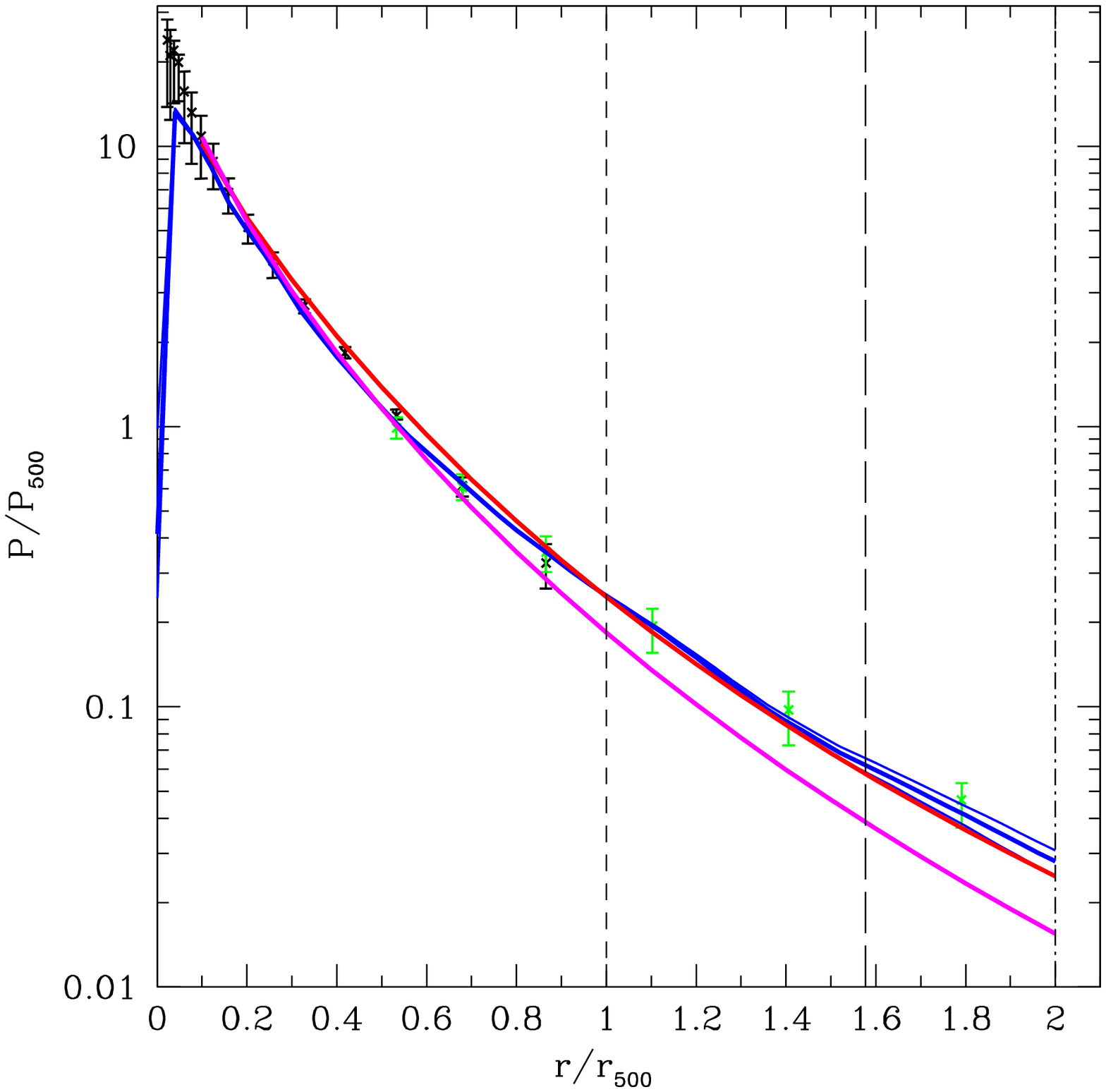}
\includegraphics[width=0.4\textwidth,height=0.3\textheight,angle=0]{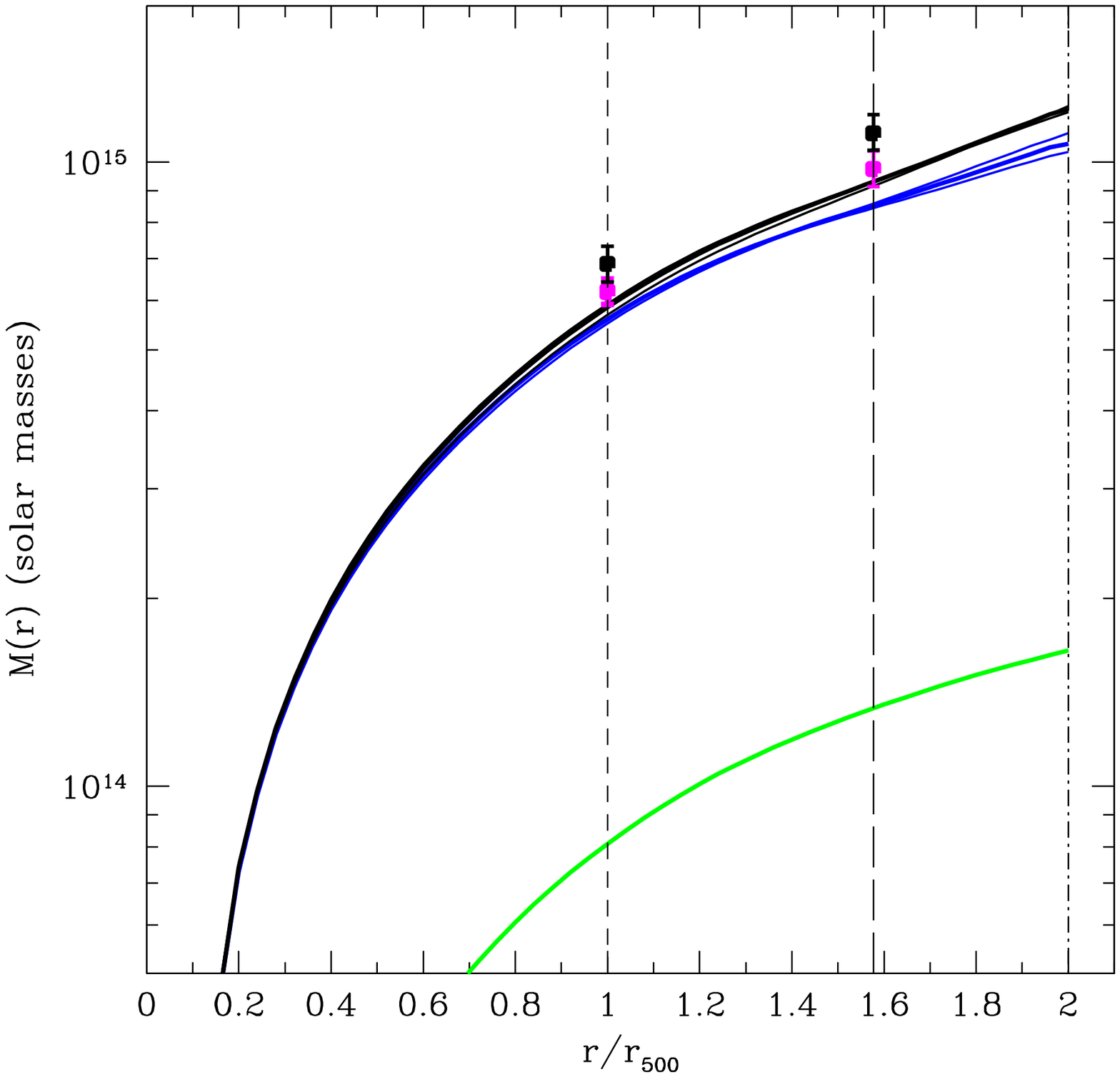}}
\caption{\textit{Top left panel}: Gas mass fraction obtained by the SM fit to the joint stacked
X-ray and SZ projected temperature profiles. The dashed area is with $\delta_R = 0$, while the thick black line is with
$\delta_R = 0.15$ and $l = 0.6$. The dashed horizontal line represents the universal baryon fraction from \textsl{Planck} (Planck Collaboration XIII 2016), whereas the thick green line with the dashed green lines is the expected gas fraction corrected for the baryon fraction in the form of stars (Gonzalez, Zaritsky \& Zabludoff 2007). \textit{Top right panel}: The black points represent the stacked entropy profile
given by \textit{XMM-Newton} observations, while the green points are derived by the \textit{Planck} survey (G18). The thick blue line is obtained by the deprojected temperature profile derived by the joint SM fit of Fig. 1 combined with the stacked gas density profile. The red line is the entropy profile under pure gravitational collapse with slope
1.1 (Voit et al. 2005). \textit{Bottom left panel}: The black points represent the stacked pressure data derived by \textit{XMM-Newton} observations, while the green points are obtained by the \textit{Planck} survey (G18). The tick blue line is given by the SM deprojected temperature profile combined with the stacked gas density profile. The red line is the pressure profile derived by the NFW functional form introduced by Nagai et al. (2007); the magenta line is the universal pressure profile of Arnaud et al. (2010). \textit{Bottom right panel}: The thick blue line is the total hydrostatic mass derived
by the radial SM temperature profile and by the stacked gas density profile (see Eq. A4 with $\delta_R = 0$); the black thick line is with $\delta_R = 0.15$ and $l = 0.6$. The magenta points are the median values derived from the mass values reported by Ettori et al. (2018) for each cluster at $r_{500}$ and $r_{200}$; the black points are the median values reported by Eckert et al. (2018) after the corrections for the presence of a nonthermal support in the cluster outskirts. The green line is the gas mass profile. In all panels, 
the thin lines represent the $1\sigma$ error, the vertical dashed, long-dashed and dot-dashed lines represent $r_{500}$, $r_{200}$, and the virial radius $R$, respectively.}
\label{fig:corr}
\end{center}
\end{figure*}


The entropy profile recovered by combining the SM deprojected temperature with the X-ray gas density through the relation $K = k_B T_{X-SZ}^{SM}/n_e^{2/3}$ is in good agreement with the stacked entropy profiles (see Fig. 2, blue line). The flattening in the central regions $k_c \simeq 140$ $\rm keV\ cm^2$ is roughly coincident with the median value derived from the analysis of the 12 X-COP clusters by Ghirardini et al. (2018). They also outline that the flattening is more evident in non-cool-clusters than in cool-core clusters as already reported in several studies (e.g., Cavagnolo et al. 2009; Pratt et al. 2010). The SM entropy profile is consistent slightly beyond $r_{500}$ with the prediction
of gravitational collapse where the entropy increases steadily out to the virial radius following a power law with slope 1.1.

Also the thermal pressure profile, $P^{SM}_{X-SZ} = k_B T_{X-SZ}^{SM} n_e$, is in good agreement with the stacked profile reported by X-ray and SZ observations, as shown by Fig. 2. The pressure profiles with
$\delta_R = 0$ and with $\delta_R > 0$ are nearly coincident for the modest level of the nonthermal pressure component derived
by the SM analysis.

The SM hydrostatic total mass profiles (see Fig, 2) with $\delta_R = 0$ and with $\delta_R > 0$ are obtained by Eq. A4.
The values of the best fit parameters are reported in Table 1, where $T_R$ is the ICM temperature at the virial radius and $c$ 
is the concentration parameter (see Cavaliere et al. 2009).

\begin{table}
	\centering
	\caption{Best fit parameters of the two SM analyses for $\delta_R = 0$.}
	\label{tab:SM_table}
	\begin{tabular}{lccc} 
		\hline
		& $T_{X-SZ}$  &  & $T_X$ \\
		\hline
		$T_R/T_{500}$ & $0.476^{+0.050}_{-0.046}$ &  & $0.128^{+0.161}_{-0.078}$ \\
		$k_c/k_R$ & $(9.4\pm 3.1)\times 10^{-3}$ &  & $(5.2\pm 2.5)\times 10^{-2}$ \\
		$a$  & $0.80\pm 0.24$ &  & $1.00\pm 0.37$ \\
		$c$ & $4.10\pm 1.12$ &  & $4.21\pm 1.15$ \\
		$r_b/r_{500}$ & -- &  & $0.40\pm 0.10$ \\
		$a^{\prime}$ & -- &  & $0.85\pm 0.45$ \\
		\hline
	\end{tabular}
\end{table}

\subsection{SM analysis of the \textit{XMM-Newton} X-ray temperature data}

The SM fit to the stacked \textit{XMM-Newton} projected temperature data requires instead an entropy profile bending outwards at
distances greater then $r_b= (0.40\pm 0.10) r_{500}$ ($a^{\prime} = 0.85\pm 0.45$), as shown by the red line in Fig. 3.
The addition of two free parameters ($r_b$ and $a^{\prime}$), with respect to the fit given by the blue line ($r_b$ = 0), is
significant at the level of 99.99\% according to the F-test ($\chi^2_{red} = 8.04/12$ for $r_b = 0$ and
$\chi^2_{red} = 0.46/10$ for $r_b > 0$). Considering that the SZ
temperature values are much less affected by gas clumping that may bias low the temperature data (Khedekar et al. 2013; Roncarelli et al. 2013; Ghirardini et al. 2018), new fits are performed replacing the outermost two X-ray
temperature points with the two SZ data at the same distances from the center obtaining a confidence level of 99.91\%
($\chi^2_{red} = 6.79/12$ for $r_b = 0$ and $\chi^2_{red} = 1.67/10$ for $r_b > 0$).
The conclusion is that these results imply that the X-ray temperature data observed
by \textit{XMM-Newton}, although limited to $r_{500}$, indicate a rapid decline of the temperature and a consequent entropy flattening
going toward the virial radius in the X-COP cluster sample. The SM extrapolation at $R$ (dashed region) is derived from the fit to the projected temperature and results in agreement with the steep profiles observed by \textit{Suzaku}, as shown by the temperature profiles of Abell 2029 (Walker et al 2012) and Abell 1795 (Bautz et al 2009), two clusters belonging to the X-COP sample (Fig.s 3 and 4). However, it should be remarked that the large error bars associated to the \textit{Suzaku} measurements in the outskirts of these clusters make these profiles not much distant from the stacked SZ temperature data. Besides, the X-COP temperature profile for Abell 2029 lies systematically below the median profile of the sample (see Fig. 5 in G18).

\begin{figure}
\includegraphics[width=\columnwidth]{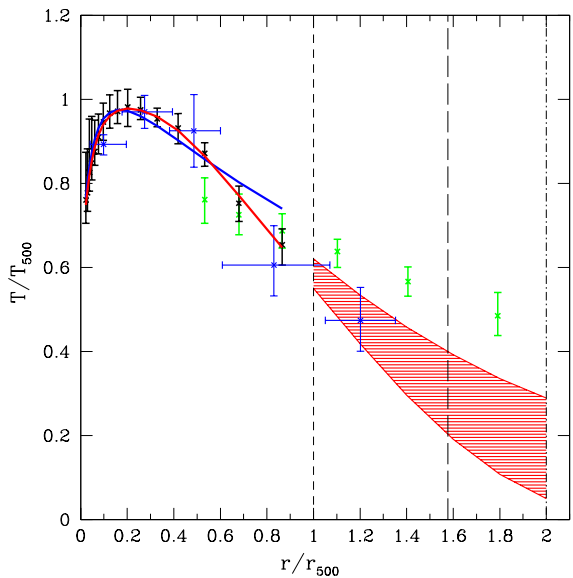}
\caption{Black and green points
represent the projected X-ray temperature data measured by \textit{XMM-Newton} and the SZ projected temperature data derived by the \textit{Planck} survey, respectively. The blue points are the temperature data reported by \textit{Suzaku} for Abell 2029 (Walker et al. 2012). The blue line is the SM fit with $r_b = 0$, while the
red line is with $r_b > 0$ (both the fits with $\delta_R = 0$). The SM fits consider only the X-ray data. The dashed area is the SM extrapolation of the fit given by the red line. The vertical dashed, long-dashed and dot-dashed lines represent $r_{500}$, $r_{200}$, and the virial radius $R$, respectively.}
\label{fig:xcop_temp_red_figure}
\end{figure}

\begin{figure}
\includegraphics[width=\columnwidth]{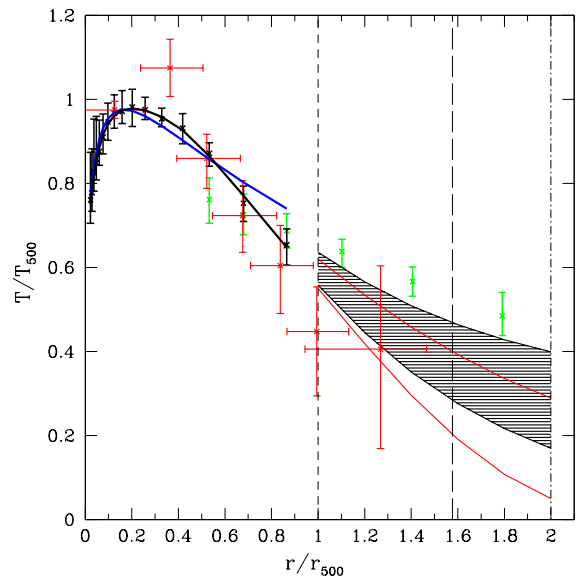}
\caption{Black and green points as in Fig. 3; red points are the temperature data reported by \textit{Suzaku} for Abell 1795 (Bautz et al 2009). The blue line as in Fig. 3; the black line is the SM fit with $r_b > 0$ and $\delta_R = 0.6$ ($l = 0.6$); the black dashed area is the SM extrapolation of the black line; the red lines are the boundaries of the red dashed area of Fig. 3. The vertical dashed, long-dashed and dot-dashed lines represent $r_{500}$, $r_{200}$, and the virial radius $R$, respectively.}
\label{fig:xcop_temp_black_figure}
\end{figure}

The radial distribution of the gas mass fraction, entropy, pressure and hydrostatic mass are recovered by combining the 3D SM temperature profile with the gas density profile corrected for clumpiness. Despite the large uncertainty associated to the temperature profile, and therefore to the total mass $M$, $f_{gas}$ results greater than the universal value at $R$ (dashed region in Fig. 5) showing the presence of a nonthermal pressure component to sustain HE in the outskirts of the X-COP clusters.
To obtain a gas mass fraction consistent with the universal value, a new fit is performed to the X-ray data with $\delta_R > 0$ in Eq. A1
to have a greater total mass (see Eq. A4). Also in this case the better fit requires an entropy profile bending outwards, as
shown by the thick black line of Fig. 4; $r_b$ and $a^{\prime}$ assume roughly the same values reported above with $\delta_R = 0$. The
dashed region is the SM extrapolation toward the virial radius of the fit with $\delta_R > 0$ to the projected temperature profile. The resulting deprojected temperature profile, $T_X^{SM}$, yields a gas mass fraction in agreement with the universal value (black lines in Fig. 5) for a nonthermal pressure component, $p_{nth}$, in the range $\sim (20-40)\%$ of the total pressure $p_{tot}$ at $R$. Eq. 1 gives $\alpha(r_{200}) \sim (18-37)\%$ and $\alpha(r_{500}) \sim (11-25)\%$.

\begin{figure*}
\begin{center}
\parbox{16cm}{
\includegraphics[width=0.4\textwidth,height=0.3\textheight,angle=0]{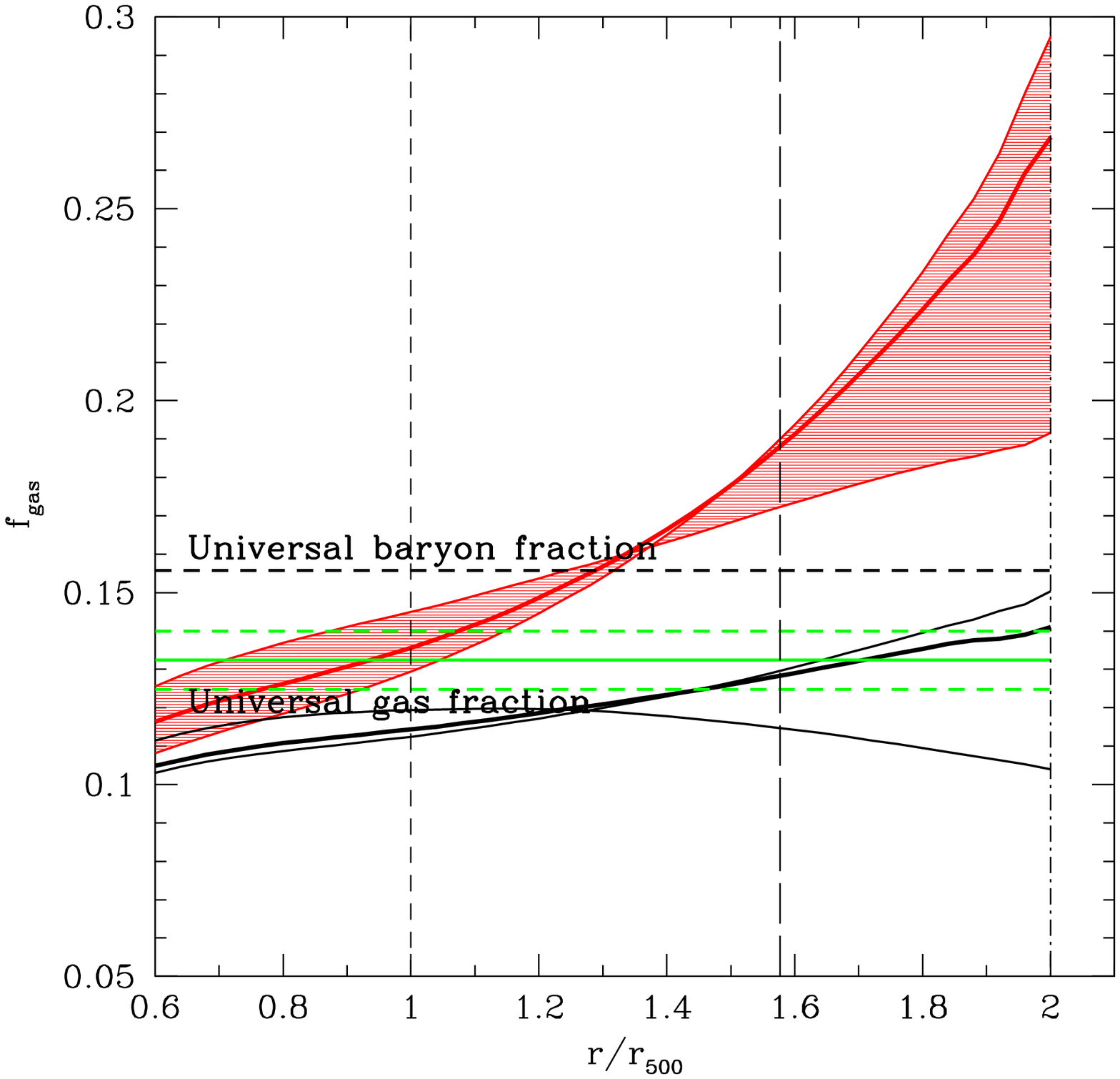}
\includegraphics[width=0.4\textwidth,height=0.3\textheight,angle=0]{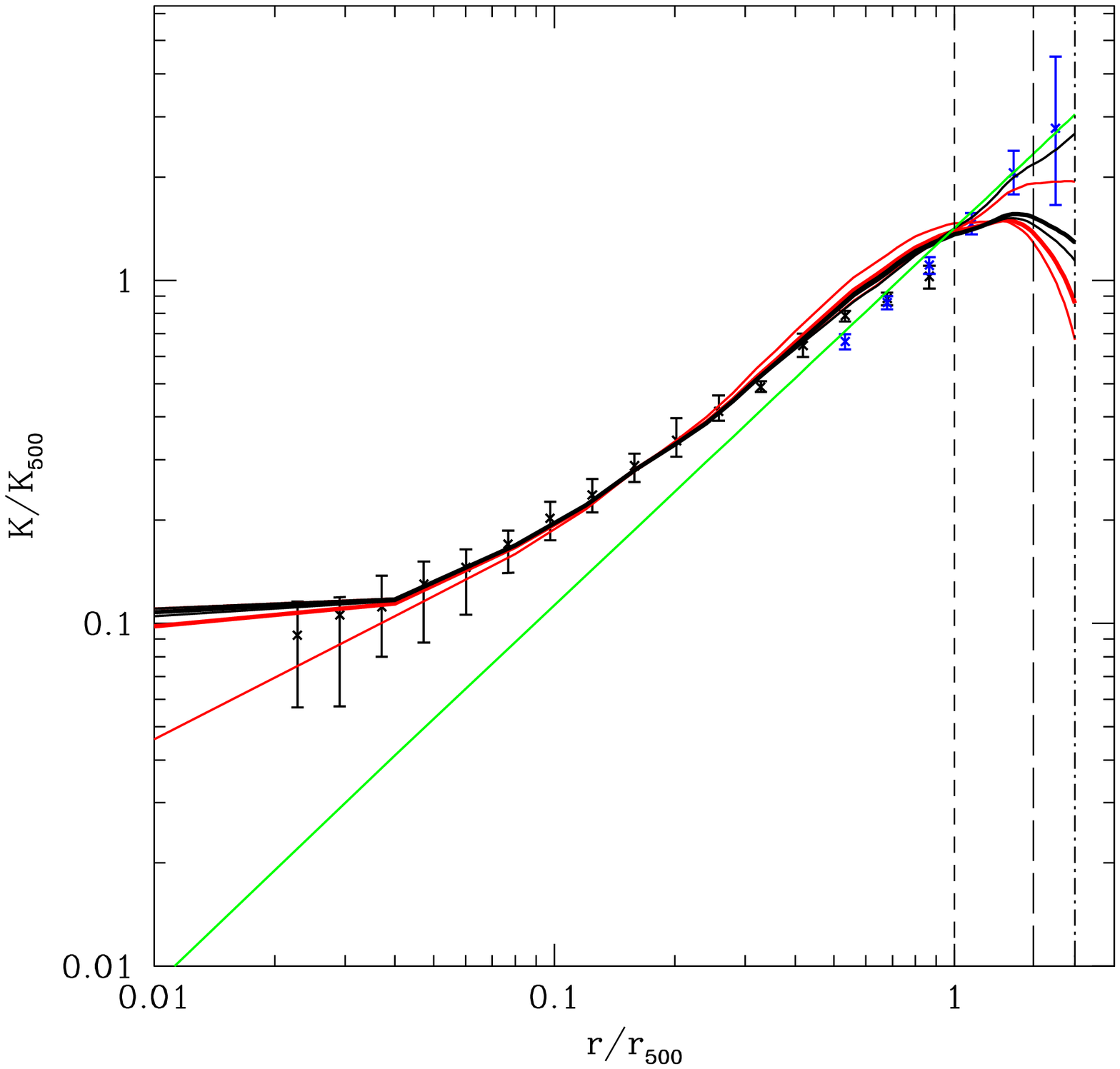}}
\parbox{16cm}{
\includegraphics[width=0.4\textwidth,height=0.3\textheight,angle=0]{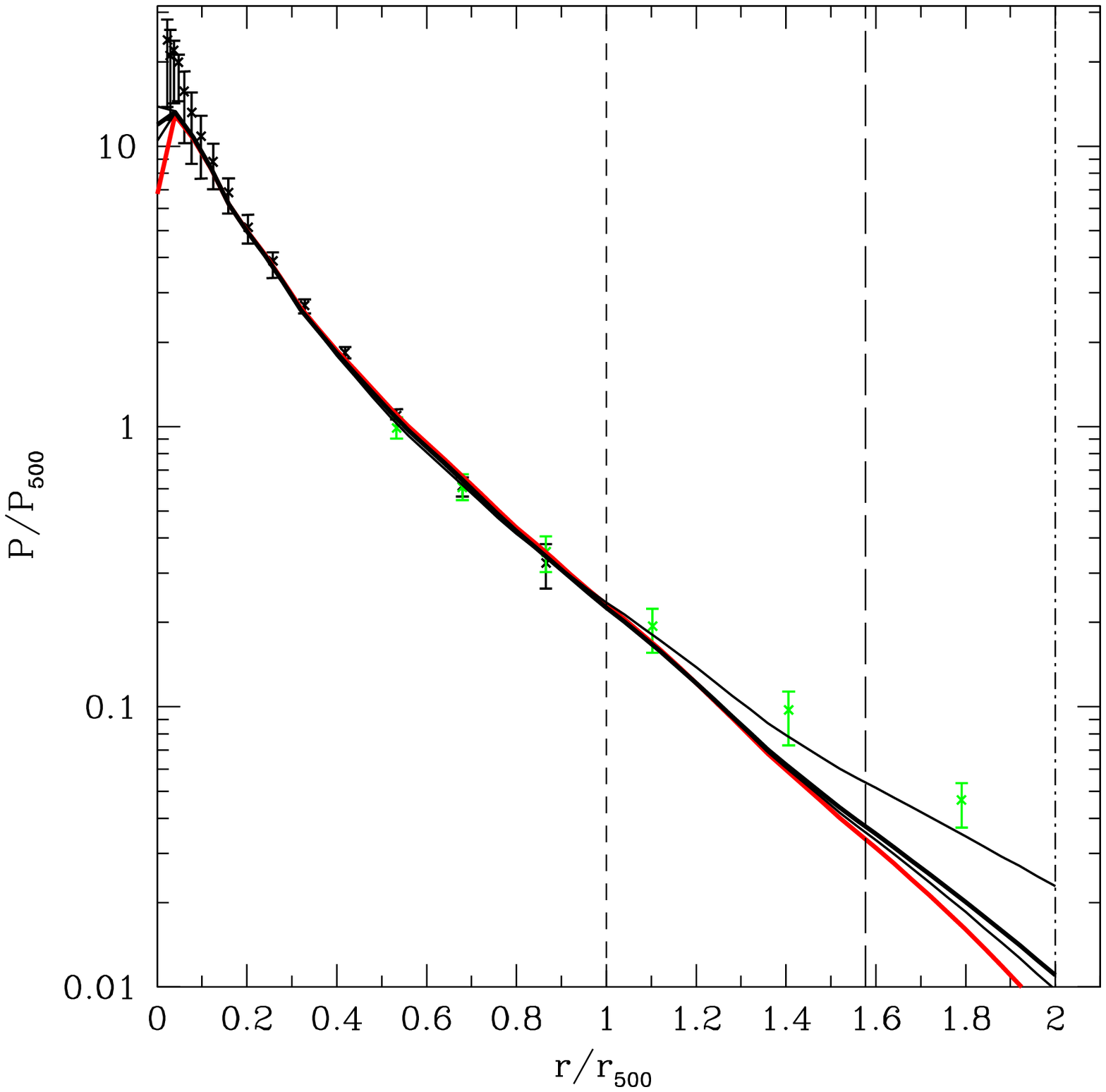}
\includegraphics[width=0.4\textwidth,height=0.3\textheight,angle=0]{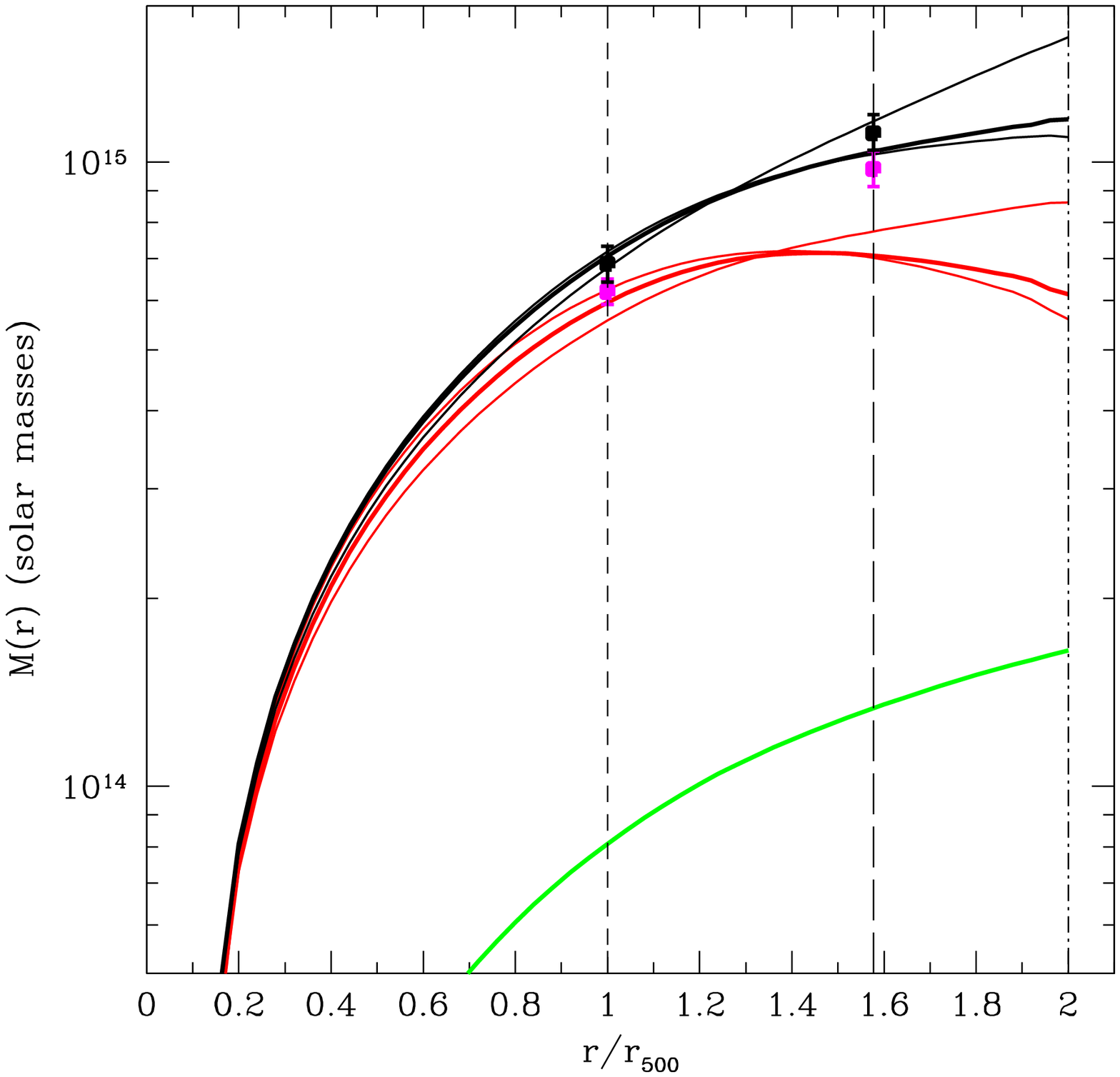}}
\caption{\textit{Top left panel}: The dashed area is the gas mass fraction derived from the SM fit to the X-ray data with $r_b > 0$ and $\delta_R = 0$ (see Fig. 3). The thick black line is given by the fit to the X-ray data (see Fig. 4) with $r_b > 0$ and $\delta_R = 0.6$ ($l = 0.6$).
For the horizontal lines see Fig. 2. \textit{Top right panel}: Black and green points are the stacked X-ray and SZ entropy data, respectively. The thick red line is the SM entropy profile obtained by the fit to the X-ray temperature data with $r_b > 0$ and $\delta_R = 0$ through the relation $K_X^{SM} = T_X^{SM}/n_e^{2/3}$; the thick black line is with $r_b > 0$ and $\delta_R = 0.6$ ($l = 0.6$). The green line is the power law increase with slope 1.1 (see text). \textit{Bottom left panel}: The black and green points are the the stacked X-ray and SZ pressure data, respectively. The red line is obtained by the SM fit to the X-ray temperature data with $r_b > 0$ and $\delta_R = 0$ through the relation $P_X^{SM} = k_BT_X^{SM}n_e$; the thick black line is with $r_b > 0$ and $\delta_R = 0.6$ ($l = 0.6$).
\textit{Bottom right panel}: The thick red line is obtained from the SM fit to the X-ray temperature data with $r_b > 0$ and $\delta_R = 0$, while the thick black line with $r_b > 0$ and $\delta_R = 0.6$ ($l = 0.6$). The green line is the gas mass. For the magenta and black points see Fig. 2. In all panels, the thin lines are the $1\sigma$ error and the vertical dashed, long-dashed and dot-dashed lines represent $r_{500}$, $r_{200}$, and the virial radius $R$, respectively.}
\label{fig:corr}
\end{center}
\end{figure*}

The SM entropy profile of the X-COP sample shows a flattening beyond $r_{500}$ (see Fig. 5) due to the rapid decline of the X-ray temperature profile, in agreement with the Suzaku observations in the outskirts of several clusters. It must be noted 
that the outermost stacked entropy point with a large error bar is only slightly above the SM entropy profile. As a consequence of the temperature decline, the SM thermal pressure profile is consistent with the X-ray pressure, but steeper than the stacked pressure profile derived by the \textit{Planck} survey (see Fig. 5).
The hydrostatic total mass profiles with $\delta_R = 0$ and with $\delta_R > 0$ are reported in Fig. 5. The behavior of the total mass profile with $\delta_R = 0$ is clearly unphysical, as discussed in the next Section.

\section{Discussion and Conclusions}

In this paper, two separate analyses of the stacked temperature data of twelve X-COP
clusters (Ettori et al. 2018; Eckert et al. 2018; Ghirardini et al. 2018) are performed via the SuperModel. In the first analysis the X-ray temperature profile observed by \textit{XMM-Newton} at distances up to $r_{500}$ and the temperature profile derived by the SZ pressure data
at distances beyond $r_{500}$ (see G18) are considered jointly. The SM fit is obtained with an entropy power law increase toward the virial radius with results that are consistent with those reported in Eckert et al. (2018) and in G18. The deprojected temperature profile with $\delta_R = 0$ combined with the stacked electron density profile, corrected for clumpiness, gives
a gas mass fraction slightly above the universal value (see Fig. 2) showing the presence of a modest nonthermal pressure
component to sustain the hydrostatic equilibrium. At $R = 2r_{500}$ the nonthermal pressure is in the range (10-15)\% of the total pressure consistent with the median values reported by Eckert et al. (2018) at $r_{500}$ and $r_{200}$ distances (see Fig. 6).
All these estimates are in disagreement with the mean nonthermal fraction of the total pressure predicted by numerical simulations (see Nelson
et al. 2014; Rasia et al., in preparation; Martizzi \& Agrusa 2016). This implies a more
efficient thermalization of the gas with respect to the predictions from the reported numerical simulations.

\begin{figure}
\includegraphics[width=\columnwidth]{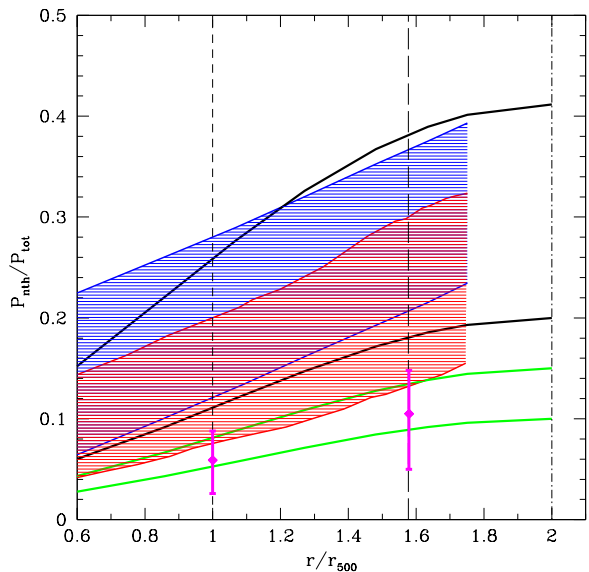}
\caption{(Adapted from Fig. 5 of Eckert et al. 2018). The blue and red lines and the shaded areas are the nonthermal pressure ratio $\alpha = p_{nth}/p_{tot}$ reported by numerical simulations of Nelson et al. (2014) and Rasia et al., respectively; the magenta points are the median values reported by Eckert et al. (2018) at $r_{500}$ and $r_{200}$. The black curves are obtained by the SM fit to the X-ray temperature data with $r_b > 0$ and $\delta_R > 0$; the green curves from the SM fit to the joint X-ray and SZ temperature data with $r_b = 0$ and $\delta_R > 0$. The vertical dashed, long-dashed and dot-dashed lines represent $r_{500}$, $r_{200}$, and the virial radius $R$, respectively.}
 \label{fig:xcop_simul_figure}
\end{figure}

To fit the thermodynamic properties Ghirardini et al. (2018) adopt piecewise power law fits where they
split the data in several radial ranges as a fraction of $r_{500}$. A second approach employs global parametric functional forms to describe the thermodynamic profiles
throughout the entire radial range. By fitting the entropy profile using piecewise power laws, G18
observe a gradual steepening of the entropy slope that becomes consistent with the prediction of gravitational collapse ($K \sim r^{1.1}$) beyond $\sim 0.5r_{500}$. On the other hand, the fit to the entropy profile throughout the entire radial range with the functional form reported in Cavagnolo et al. (2009), which consists of a power law with a constant entropy floor, does not report a slope compatible with the data that appears instead consistent with the self-similar prediction of 1.1 beyond $0.6r_{500}$. The SM deprojected temperature profile,
$T_{X-SZ}^{SM}$, combined with
the stacked electron density profile, gives an entropy profile that follows very well the
stacked entropy profile and the power law increase with slope 1.1 at distances slightly above
$r_{500}$ (see Fig. 2). The entropy floor $K_c \sim 137$ $\rm keV\ cm^2$ is
in remarkable agreement with the median value
of 140 $\rm keV\ cm^2$ reported by G18. This value reflects the preponderant presence of non-cool-core
clusters in the sample flattening more than cool-core-clusters (e.g., Cavagnolo et al. 2009; Pratt et al. 2010).

The stacked pressure profile reported in G18 is obtained by combining the X-ray pressure ($P_X = k_B T_X n_e$) in the inner regions
of the X-COP clusters with the direct deprojection of the SZ effect beyond $r_{500}$. This profile is within the envelopes
obtained by the \textit{Planck} collaboration for a sample of 62 clusters (\textit{Planck} collaboration et al. 2013). The SM pressure profile
$P_{X-SZ}^{SM} = k_B T_{X-SZ}^{SM} n_e$ is in good agreement with the stacked profile (see Fig. 2, blue line). The figure reports also
the fit of G18 to the pressure profiles of the entire population of clusters using the generalized NFW functional form introduced by Nagai et al. (2007) and the universal pressure profile of Arnaud et al. (2010). The latter lies below the stacked pressure profile
toward the virial radius (see Fig. 4) consistent with the prevalence of merging clusters in the X-COP sample. The
observed pressure profiles of non-merging clusters are in fairly good agreement with the universal profile, unlike the
profiles of merging clusters that tend to lie above (e.g., Walker et al. 2018).

Ettori et al. (2018) applying two methods have derived the hydrostatic mass of the twelve massive galaxy clusters that has been compared with independent estimates. In the \textit{backward} method they predict a temperature profile, obtained by an assumed parametric mass model and by the gas density profile, that when compared to the measured profile by X-ray and SZ observations allows to constrain the mass model parameters. Among the various mass models that have been considered they adopt the NFW mass model (Navarro et al. 1996) as reference model. Systematic uncertainties may be introduced that depend on the shape of the model and on the limited number of two
parameters of the assumed model. Besides, others uncertainties may be given by the violation of the assumed sphericity of the gas
distribution (Sereno et al 2017) and by the absence of a nonthermal pressure component to sustain the hydrostatic equilibrium. This component has been evaluated by Eckert et al. (2018) with a median value of $\sim 6\%$ at $r_{500}$ and $\sim 10\%$ at $r_{200}$ of the
total pressure in good agreement with the SM analysis, as reported above. The SM stacked mass profile in Fig. 2
is consistent at $r_{500}$ with the median value reported by Ettori et al. (2018)
for the X-COP sample but slightly lower than the median value at $r_{200}$. The median mass values with nonthermal corrections applied by Eckert et al. (2018) are above the SM mass profile with $\delta_R > 0$ at both the distances.

The second SM analysis regards only the X-ray temperature profile given by \textit{XMM-Newton} observations limited to $r_{500}$. A similar SM analysis has been performed for Abell 2142 (Fusco-Femiano \& Lapi 2018) that reports $p_{nth} \sim 30\%$ of the total pressure at the virial radius $R = 2r_{500}$, at variance with the analysis of Tchernin et al. (2016) that find only a modest presence of a nonthermal support. Their conclusion was that when gas clumping is considered, radial profiles of the gas mass fraction, entropy and hydrostatic mass
are in agreement with predictions. The discrepancy between the two analysis is explained in Eckert et al. (2018) by the presence of accreting substructures which bias low the measured \textit{XMM-Newton} temperature in the outermost region around $r_{500}$ and therefore
the SM extrapolation underestimates the \textit{Planck} data beyond $r_{500}$.
This claim is not supported by the SM mass profile derived for Abell 2142 that results consistent with \textit{all} the measurements
reported in the literature not only at $r_{500}$ but also at $r_{200}$, in particular with the $M_{200}$ value from Subaru weak lensing (Umetsu et al. 2009) as shown by Fig. 4 of Fusco-Femiano \& Lapi (2018). This implies the reliability of the SM temperature profile
also toward the virial radius. Besides, the presence of inhomogeneities
in the accreting matter is relevant for the X-ray flux at distances greater than $r_{500}$ as shown by simulations (Vazza et al. 2013; Roncarelli et al. 2013; Eckert et al. 2015), and, as reported in several points of the paper of Tchernin et al. (2016), the temperature profile from
spectroscopic analysis is mildly affected by the presence of clumps. To reinforce this conclusion it is important to stress that they derive a clumping factor $\sqrt C \simeq 1$ around $r_{500}$. Consistently with our findings Eckert et al. (2017), basing on X-ray Chandra observations of Abell 2142, report that infall of groups can generate turbulence.
Recently, Eckert et al. (2018) in their analysis of the X-COP
clusters obtain for Abell 2142 $\alpha (r_{200}) = 18.6^{+7.1}_{-8.8}\%$ and $\alpha (r_{500}) = 15.8^{+4.5}_{-4.8}\%$, only slightly different from the values derived by Fusco-Femiano \& Lapi (2018) of $\sim 25\%$ and $\sim 10\%$ at $r_{200}$ and $r_{500}$, respectively.

The relevant point of this second analysis is that the SM fit to the \textit{XMM-Newton} X-ray temperature data requires an entropy profile
bending outwards at distances $r > r_b$ that implies a steep decline of the temperature toward the virial radius. It is important
to stress that the outermost two X-ray temperature points are consistent with the SZ temperature data at the same distances from the center (see Fig. 1), as confirmed by the fits when the outermost two X-ray temperature points are replaced with the SZ
values (see Section 2.2). This implies that the results of the second SM analysis do not depend on the presence of gas clumpiness considering that the SZ temperature data are much less affected by gas clumping (e.g. Khedekar et al. 2013; Roncarelli et al. 2013; G18). In agreement with this, a negligible effect by clumping is reported by simulations on the X-ray flux measured up to $r_{500}$.
The SM temperature extrapolation toward
the virial radius is consistent with the findings of the \textit{Suzaku} observations as shown by the temperature profiles of two X-COP clusters: Abell 2029 and Abell 1795 reported in Fig.s 3 and 4.

The rapid decline of the temperature in the cluster outskirts leads to a
gas mass fraction well above the universal value (see Fig. 5) even if the gas density profile corrected by accreting structures and inhomogeneities gives a lower gas mass $M_{gas}$. The $f_{gas}$ profile highlights the presence of a nonthermal pressure component for the HE with $\alpha (R)$ in the range $(20-40)\%$, in good agreement with the numerical simulations of Nelson et al. (2014) and Rasia et al. (see Fig. 6) and with the simulations of Martizzi \& Agrusa (2016). The latter authors show a nonthermal pressure support that, though negligible in the inner regions, is increasingly relevant toward the outskirts, yielding a contribution of $(20-40)\%$ to the total pressure at radii in the range $r\sim r_{500} - 2\, r_{500}$. This contribution is considered a lower limit to the pressure balance in clusters because additional sources as magnetic fields and cosmic rays are currently not included in the simulations.

The entropy profile $K^{SM}_X = T^{SM}_X/n_e^{2/3}$ shows a flattening beyond $r_{500}$ (see Fig. 5) in agreement with several \textit{Suzaku} observations. Furthermore, for massive clusters ($M_{500} > 2 \times 10^{14}\, M_{\odot}$), as the clusters of the
X-COP sample, these observations report scaled temperatures below the expectations from simulations,
leading to an accentuated entropy flattening, unlike for the low mass clusters and merging systems outside $r_ {200}$ (Walker et al. 2013; Walker et al. 2018).
Considering that the $n_e$ profile is corrected for clumpiness, the entropy flattening is due to the steep decline of the temperature in agreement with the conclusions of Okabe et al. (2014) in their joint X-ray and weak lensing study of four relaxed galaxy clusters observed by \textit{Suzaku} and \textit{Subaru} out to virial radii.

The SM pressure profile is in good agreement with the X-ray profile but it is below the SZ data toward the
virial radius for the rapid decline of the SM deprojected temperature profile. As reported in several \textit{Suzaku} observations
the steeply decreasing temperature profile gives an unphysical total mass profile that requires the presence of a nonthermal pressure component for the hydrostatic equilibrium. The SM mass profile with $\alpha (R) \sim (20-40)\%$ is in good agreement with the median values derived by Eckert et al. (2018) after the non-gravitational corrections to the mass determinations of Ettori et al (2018) (see Fig. 5).

G18 found beyond $\sim 0.5 r_{500}$ a temperature slope of $-0.4$ in their joint X-SZ analysis that appears much flatter than the
typical slope of $-1$ observed for a dozen clusters by \textit{Suzaku} (Reiprich et al. 2013). The authors attribute this discrepancy to the low angular resolution of \textit{Suzaku} that does not allow to remove cool and overdense regions that could bias low the measured spectroscopic temperature. This statement is not supported by the SM analysis of only the \textit{XMM-Newton} X-ray data
that reports a temperature profile toward the virial radius consistent with the steep \textit{Suzaku} profiles. It is
important to stress that the \textit{XMM-Newton} observations are limited to $r_{500}$ where the presence of inhomogeneities is expected to be irrelevant for the X-ray flux, as discussed above. As well, the flatter SZ stacked temperature profile with respect to the steep \textit{Suzaku} profiles cannot be explained by the prevalence of NCC clusters in the X-COP cluster sample as evidenced by the analysis of G18. They found steeper temperature profiles in the outskirts of NCC clusters rather than in CC systems, with a marginally significant entropy flattening of the NCC population.

Some progress has been made in recent years in our understanding of the cluster outskirts. However, the low density and temperature of the gas in these external regions make X-ray and SZ measurements highly challenging. It is beyond the scope of this work to
discuss the goodness of the temperature profile measured by \textit{Suzaku} or \textit{Planck} observations. The main result reported here concerns the SM analysis of the \textit{XMM-Newton} X-ray data that shows that a steep temperature profile may be present in the X-COP cluster outskirts instead of the flatter profile reported by the \textit{Planck} survey. In addition, the two analyses have evidenced different ICM thermodynamic properties.
The SZ temperature profile involves a modest presence of a nonthermal pressure component, at variance with numerical simulations, and an entropy profile that follows beyond $\sim (0.5-1)r_{500}$ the predicted power law increase with slope $1.1$. Conversely, the rapid decline of the temperature reported by the SM analysis of the X-ray data, in good agreement with the \textit{Suzaku} observations, implies a more relevant level of the nonthermal pressure support consistent with the values derived by numerical simulations, and an entropy flattening beyond $r_{500}$.

Planned X-ray missions and upcoming SZ experiments and instrumentations will give maps
of the thermodynamical properties and will reveal the level of the nonthermal support in the cluster outskirts (e.g., Walker et al. 2018).

\section*{Acknowledgements}
I am grateful to P. Mazzotta for stimulating discussions. I thank the referee for valuable comments.

\appendix

\section{SM equations}

In the presence of turbulence a nonthermal pressure component is added to the thermal one giving a total pressure
$p_{\rm tot}(r) = p_{\rm th}(r) + p_{\rm nth}(r) = p_{\rm th}(r)[1 +
\delta(r)]$ where $\delta(r) \equiv p_{\rm nth}/p_{\rm
th}$ that when inserted in the HE equation yields the temperature profile as
\begin{equation*}
\frac{T(r)}{T_R} = \left[\frac{k(r)}{k_R}\right]^{3/5}\, \left[\frac{1 +
\delta_R}{1 + \delta(r)}\right]^{2/5}\,\times
\end{equation*}
\begin{equation}
\left\{1 + \frac{2}{5}\frac{b_R}{1 +
\delta_R}\int_r^R {\frac{{\rm d}x}{x} \frac{v^2_c(x)}{v^2_R}\,
\left[\frac{k_R}{k(x)}\right]^{3/5}\, \left[\frac{1 + \delta_R}{1 +
\delta(x)}\right]^{3/5}}\right\}
\end{equation}

\par\noindent
and the pressure profile as
\begin{equation*}
\frac{P(r)}{P_R} = \left[\frac{1 + \delta_R}{1 + \delta(r)}\right]\,\times
\end{equation*}
\begin{equation}
\left\{1 + \frac{2}{5} \frac{b_R}{1 + \delta_R}\,
\int_r^R {\frac{{\rm d}x}{x}
\frac{v^2_c(x)}{v^2_R} \left[\frac{k_R}{k(x)}\right]^{3/5}\, \left[\frac{1 +
\delta_R}{1 + \delta(x)}\right]^{3/5}}\right\}^{5/2}
\end{equation}

\par\noindent
where $v_c$ is the DM circular velocity ($v_R$ is the value at the virial radius $R$), and $b_R$
is the ratio at $R$ of $v^2_c$ to the sound speed squared (Cavaliere et al. 2009; Cavaliere et al. 2011).
The functional shape, $\delta(r)$, is given by Eq. 1.

For $k(r)$ two profiles are adopted for the SM analyses: \textit{i)} a spherically averaged profile with shape
$k(r) = k_c + (k_R - k_c)(r/R)^a$ (see Voit 2005); \textit{ii)} an entropy run that starts as a
powerlaw with slope $a$, but it flattens for radii $r > r_b$ (Lapi et al. 2010). For the sake of simplicity, a 
linear entropy decline is taken with gradient $a^{\prime} \equiv (a-a_R)/(R/r_b - 1)$,
where $r_b$ and $a^{\prime}$ are free parameters to be determined from the
fitting of the ICM temperature profile. This entropy profile allows to model the entropy flattening with respect to a
pure gravitational inflow and the steep temperature profiles reported by several X-ray observations.


The traditional equation to estimate the total X-ray mass $M(r)$ within $r$
is modified as follows in presence of an additional nonthermal
pressure component (Fusco-Femiano \& Lapi 2013)

\begin{equation*}
M(r) = - \frac{k_B [T(r)(1 +\delta(r)] r^2 }{\mu m_p G}\,\times
\end{equation*}
\begin{equation*}
\left\{\frac{1}{n_e(r)}\frac{d n_e(r)}{d r} +
\frac{1}{T(r)[(1+\delta(r)]}\frac{d T(r)[1 +
\delta(r)]}{d r}\right\}
\end{equation*}
\begin{equation*}
= - \frac{k_B [T(r)(1 +\delta(r)] r^2}{\mu m_p G}\,\times
\end{equation*}
\begin{equation}
\left[\frac{1}{n_e(r)}\frac{d n_e(r)}{d r} +
\frac{1}{T(r)}\frac{d T(r)}{d r} + \frac{\delta(r)}{1
+ \delta(r)} \frac{2}{l^2}(R - r)\right]~.
\end{equation}
The gas mass is given by
$$M_{\rm gas} = 4\pi \mu_e m_p\int{\rm d}r~{n_e(r) r^2}$$
where $\mu_e \sim 1.16$ is the mean molecular weight of the electrons.

\bsp	
\label{lastpage}
\end{document}